\documentclass[superscriptaddress, reprint, amsmath, amssymb, aps, pra, floatfix]{revtex4-2}

\usepackage[utf8]{inputenc}
\usepackage{graphicx}% Include figure files
\usepackage{dcolumn}% Align table columns on decimal point
\usepackage{bm}% bold math
\usepackage[normalem]{ulem} % For strikethrough
\usepackage{mathrsfs}
\usepackage{physics}
\usepackage{xcolor}
\usepackage{amsmath, amssymb, mathtools, amsthm}
\usepackage[colorlinks=true]{hyperref}
\usepackage{comment} %create block comment
\usepackage{orcidlink}

\usepackage{xr}
\graphicspath{{Figures/}} %figures path

\definecolor{ve}{RGB}{0,161,22}

\begin{abstract}
Sublinear optical response, in which the emitted field scales as a fractional power of the driving field, lies beyond the conventional perturbative hierarchy of nonlinear optics. Here, we show that such a response can be engineered in a hydrogen atom using tracking control. Rather than prescribing the driving waveform, the field is determined self-consistently from the evolving quantum state to enforce a chosen relation between the optical response and the applied field. We demonstrate accurate tracking for multiple exponents and scaling strengths, establishing that a single atomic system can be driven to realize a family of distinct sublinear responses. The calculations are enabled by a compact wave-packet continuum discretization that treats bound and continuum states on equal footing and is applied here, to the best of our knowledge, for the first time to strong-field optics and quantum control. These results establish tracking control as a general route to engineering optical responses beyond conventional polynomial nonlinearities.

\end{abstract}

\begin{document}

\author{Mykhaylo Khoma}
\email{mykhaylo.khoma@matfyz.cuni.cz}
\affiliation{Institute of Theoretical Physics, Faculty of Mathematics and Physics, Charles University, V Hole\u{s}ovi\u{c}k\'{a}ch 2, 18000 Prague, Czech Republic}

\author{Valeriia Bilokon\orcidlink{0009-0001-1891-0171}}
%\thanks{These authors contributed equally}
\email{vbilokon@tulane.edu}
\affiliation{Department of Physics and Engineering Physics, Tulane University, New Orleans, Louisiana 70118, United States}
\affiliation{Akhiezer Institute for Theoretical Physics, NSC KIPT, Akademichna 1, 61108 Kharkiv, Ukraine}

\author{Elvira Bilokon\orcidlink{0009-0007-8296-2906}}
%\thanks{These authors contributed equally}%\footnotemark[\value{footnote}] 
\email{ebilokon@tulane.edu}
\affiliation{Department of Physics and Engineering Physics, Tulane University, New Orleans, Louisiana 70118, United States}
\affiliation{Akhiezer Institute for Theoretical Physics, NSC KIPT, Akademichna 1, 61108 Kharkiv, Ukraine}

\author{Denys I. Bondar~\orcidlink{0000-0002-3626-4804}}
\email{dbondar@tulane.edu}
\affiliation{Department of Physics and Engineering Physics, Tulane University, New Orleans, Louisiana 70118, United States}

\title{Nonlinear Response via Sublinear Optics}

\maketitle

%-----------------------------------------------------
%%%%% Introduction %%%%%
%-----------------------------------------------------
\section{Introduction}
The perturbative expansion provides the foundation of nonlinear optics and organizes a vast range of light--matter phenomena~\cite{Bloembergen1996, Shen1984, Boyd2008}. Within this framework, the optical response is expressed as a series of integer-order contributions in the driving field~\cite{Armstrong1962}. This classification extends from familiar effects such as second-harmonic generation~\cite{Franken1961} to high-harmonic generation, where effective nonlinear orders exceeding one hundred have been observed~\cite{McPherson1987, Ferray1988, Seres2005, Krausz2009}. Despite its broad success, the perturbative description assumes that the response is analytic in the field and therefore restricts the allowed scaling to integer powers. Optical responses that grow more slowly than the driving field lie outside this conventional hierarchy and remain largely unexplored.

This limitation has helped drive a broader shift in nonlinear optics: from characterizing the response of naturally occurring materials to engineering desired optical properties by design. Over the past two decades, metamaterials, metasurfaces, and photonic crystals have enabled such control through tailored geometry and composition, producing effective susceptibilities unavailable in conventional media~\cite{Kuznetsov2024, Li2017, Koshelev2023, Schulz2024, Zograf2024, Wang2024, Park2024}. Epsilon-near-zero materials, whose permittivity approaches zero near a characteristic frequency, have attracted particular attention because of their unusually strong nonlinear response~\cite{Alam2016, Reshef2017, Niu2018, Reshef2019, Yang2019, Fomra2024, Wu2024}. In all these approaches, however, the design freedom remains encoded in the material structure, while the driving field is treated as a prescribed input. The resulting responses, whether enhanced or suppressed, still belong to the conventional hierarchy of integer-order susceptibilities.

An alternative route is to shift the design freedom from the material to the driving field. Rather than fabricating a medium with a desired response, one asks which time-dependent field will make a given quantum system follow a prescribed optical trajectory. This is the central idea of tracking control~\cite{Gross1993, Zhu1999, Koch2022}. Its development has established that suitably shaped fields can reproduce target observables across a broad range of quantum systems~\cite{Campos2017,Magann2018,McCaul2020PRL,McCaul2020PRA, McCaul2021, Masur2023}. More recently, advances in waveform synthesis have made increasingly complex control fields experimentally accessible~\cite{Alqattan2022, McCaul2023PRL, Peng2025, McCaul2025}.

Despite this progress, optical responses that grow more slowly than the driving field remain largely unexplored. Such sublinear behavior cannot be captured by the conventional expansion in integer-order susceptibilities and therefore represents a distinct form of nonperturbative optical response. In this work, we prove that a quantum system can be driven to exhibit prescribed sublinear behavior through quantum tracking control. This is also illustrated computationally for the paradigmatic example of a hydrogen atom in 3D. Moreover, we achieve an order-of-magnitude speedup and a reduction in the memory overhead of the required quantum dynamical simulations by employing a wave-packet continuum discretization basis~\cite{Rubtsova2015,Pomerantsev2016} for our control calculations, which treats bound and continuum states within a compact representation. To the best of our knowledge, this is the first application of this approach to strong-field optics and quantum control. We demonstrate robust tracking across a range of sublinear responses, establishing a practical route to optical behaviors beyond the usual perturbative hierarchy. Sublinear response, which by construction enhances contrast at low field amplitudes, may find application in optical sensing and dynamic-range compression, and more broadly extends the landscape of engineerable light--matter interactions.

%-----------------------------------------------------
%%%%% Model %%%%%
%-----------------------------------------------------
\section{Model and Methods}
The Hamiltonian describing the electronic motion in a hydrogen atom exposed to an external field is taken in the form
\begin{equation}\label{eq:H_general}
  H(\mathbf{r}, t) = T + V_0(\mathbf{r}) + V_f(\mathbf{r}, t),
\end{equation}
where $\mathbf{r}$ is the position vector of the electron,
$T$ is the kinetic energy operator of the electron, $V_0(\mathbf{r})$ is the Coulomb potential of the parent ion, and $V_f(\mathbf{r}, t) = \mathbf{r} \cdot \bm{\mathcal{E}}(t)$ couples the electron to the applied field $\bm{\mathcal{E}}(t)$ within the dipole approximation. We consider a linearly polarized laser field $\bm{\mathcal{E}}(t) = \mathcal{E}(t)\,\hat{\mathbf{z}}$, so that $V_f = z\, \mathcal{E}(t)$. Throughout, we use atomic units $\hbar = m_e = e = 1$, with the convention that the electron has charge $-e$.

%-----------------------------------------------------
%%%%% Wave-Packet Continuum Discretization Approach %%%%%
%-----------------------------------------------------
\subsection{Wave-Packet Continuum Discretization Approach}
To represent the field-free Hamiltonian $H_0 = T + V_0$ and the relevant operators in a finite basis that includes continuum states on equal footing with bound states, we employ the wave-packet continuum discretization (WPCD) approach~\cite{Rubtsova2015, Pomerantsev2016}. The WPCD formalism has been successfully applied to a variety of problems in atomic and nuclear collision physics~\cite{Pomerantsev2016, Abdurakhmanov2016, Miller2022}. A review of the construction of continuum-discretized states is given in Ref.~\cite{Rubtsova2015}. Here, we provide only a brief summary of the method.

The central idea, originating from the work of Weyl~\cite{Weyl1910}, Bethe~\cite{Bethe1933}, and Wigner~\cite{Wigner1959}, is to replace pure continuum states---which are not normalizable and hence do not belong to the Hilbert space---by wave packets obtained by averaging them over narrow energy or momentum intervals. The resulting functions are square-integrable and can therefore be treated within the standard Hilbert-space formulation of Hermitian quantum mechanics.

To discretize the continuum states of the field-free atomic Hamiltonian $H_0$, we follow the standard procedure described in Refs.~\cite{Rubtsova2015, Abdurakhmanov2016}. We consider a finite energy interval $[E_0,E_{\max}]$ that is sufficient to describe the dynamics of the system under consideration. This interval is divided into $N$ non-overlapping energy bins $[E_{i-1},E_i]$ of width $\Delta E_i = E_i - E_{i-1}$, with $i=1,2,\ldots,N$ and $E_N \equiv E_{\max}$. 

For each energy bin, the corresponding wave-packet state is constructed by averaging the continuum eigenstates $\psi^{\rm c}_\ell$ over that interval,
\begin{equation}
\psi^{\rm wp}_{i\ell}(\Delta E_i)
=
\frac{1}{\sqrt{\Delta E_i}}
\int_{E_{i-1}}^{E_i}
\psi^{\rm c}_\ell(E')\,dE',
\end{equation}
where $\ell$ is the angular-momentum quantum number and $\langle \psi^{\rm c}_\ell(E) | \psi^{\rm c}_\ell(E') \rangle = \delta(E-E')$.
For fixed $\ell$, the wave-packet states form an orthonormal discrete representation of the continuum,
\begin{equation}
\langle \psi^{\rm wp}_{i\ell} | \psi^{\rm wp}_{j\ell} \rangle
=
\delta_{ij}.
\end{equation}
Each state is associated with the mean energy of its corresponding bin,
$e_i = (E_{i-1}+E_i)/2$, so that $H_0$ is diagonal in the wave-packet basis.

The selected bound states $\psi^{\rm b}_{n\ell}$ of the same $H_0$, together with the continuum wave packets $\psi^{\rm wp}_{i\ell}$, form an orthonormal finite basis. The wave function of the full Hamiltonian can then be expanded as
\begin{equation}
|\Psi\rangle =
\sum_{n, \ell} c_n |\psi^{\rm b}_{n\ell}\rangle
+
\sum_{i,\ell} c_{i\ell}\,|\psi^{\mathrm{wp}}_{i\ell}\rangle,
\end{equation}
where both sums run over $\ell \leq L_{\max}$, with $n \leq n_{\max}$ for the bound states and $i = 1,\ldots,N$ for the continuum bins.
This basis provides a practical finite-dimensional representation in which bound and continuum contributions enter on the same footing. In the present work we retain bound states up to $n_{\max}=17$ ($L_{\max}=6$ for $n \leq$ 16 and $L_{\max}=0$ for $n=17$) and discretize the continuum into 60 energy bins per angular momentum, yielding a 512-state basis (with 92 bound and 420 continuum states).

%-----------------------------------------------------
%%%%% Tracking Control %%%%%
%-----------------------------------------------------
\subsection{Tracking Control}
According to the Larmor formula, the field radiated by a charged particle is proportional to its acceleration.  For a single active electron in the dipole approximation, the relevant acceleration is $R(t) \equiv d\langle p_z\rangle/dt$, which we adopt as the optical response throughout this work. The goal of tracking control is to engineer a prescribed functional relationship between the response $R(t)$ and the driving field $\mathcal{E}(t)$. In this study, we target a sublinear power-law response
\begin{equation}
  R_{\rm target}(t) = \alpha \operatorname{sgn}(\mathcal{E}) |\mathcal{E}(t)|^{1/\nu},
  \label{eq:target_response}
\end{equation}
where $\alpha$ is a constant and $\nu > 1$ parametrizes the degree of sublinearity. This response cannot be expressed through the standard perturbative expansion in powers of $\mathcal{E}$ and represents a qualitatively distinct class of optical nonlinearity.

The Ehrenfest theorem applied to Hamiltonian~\eqref{eq:H_general} states
\begin{equation}
    R(t) = \frac{d}{dt}\langle p_z \rangle 
    = \langle\Psi(t)| F_z |\Psi(t)\rangle - \mathcal{E}(t),
    \label{eq:ehrenfest}
\end{equation}
where $F_z = -\partial V_{0} /\partial z.$ The tracking control ensures $R(t) = R_{\rm target}(t)$. Hence, substituting Eq.~\eqref{eq:target_response} into Eq.~\eqref{eq:ehrenfest} and solving for $\mathcal{E}(t)$ yields the tracking equation,
\begin{equation}
 \alpha \operatorname{sgn}(\mathcal{E}) |\mathcal{E}(t)|^{1/\nu}
  + \mathcal{E}(t) 
  = \langle\Psi(t)| F_z |\Psi(t)\rangle,
  \label{eq:tracking}
\end{equation}
a nonlinear algebraic equation that must be solved at each moment of time $t$: the right-hand side depends on the quantum state $|\Psi(t)\rangle$, while the solution $\mathcal{E}(t)$ determines the field that drives that state. Note that a solution of Eq.~\eqref{eq:tracking} always exists. 

Equation~\eqref{eq:tracking} is solved numerically via a time-stepping procedure. At each step: (i)~the expectation value $\langle F_z\rangle$ is computed from the current state $|\Psi(t)\rangle$; (ii)~the tracking equation~\eqref{eq:tracking} is solved for the field $\mathcal{E}(t)$; (iii)~the state $|\Psi\rangle$ is propagated forward by one time step under $H(t) = H_0 + z\,\mathcal{E}(t)$. We note that evaluating $\langle F_z\rangle$ at every time step requires the full Coulomb force matrix, including continuum contributions; this is precisely what the WPCD basis provides.

%-----------------------------------------------------
%%%%% Results %%%%%
%-----------------------------------------------------
\section{Results}

\begin{figure}
    \centering
    \includegraphics[width=0.85\linewidth]{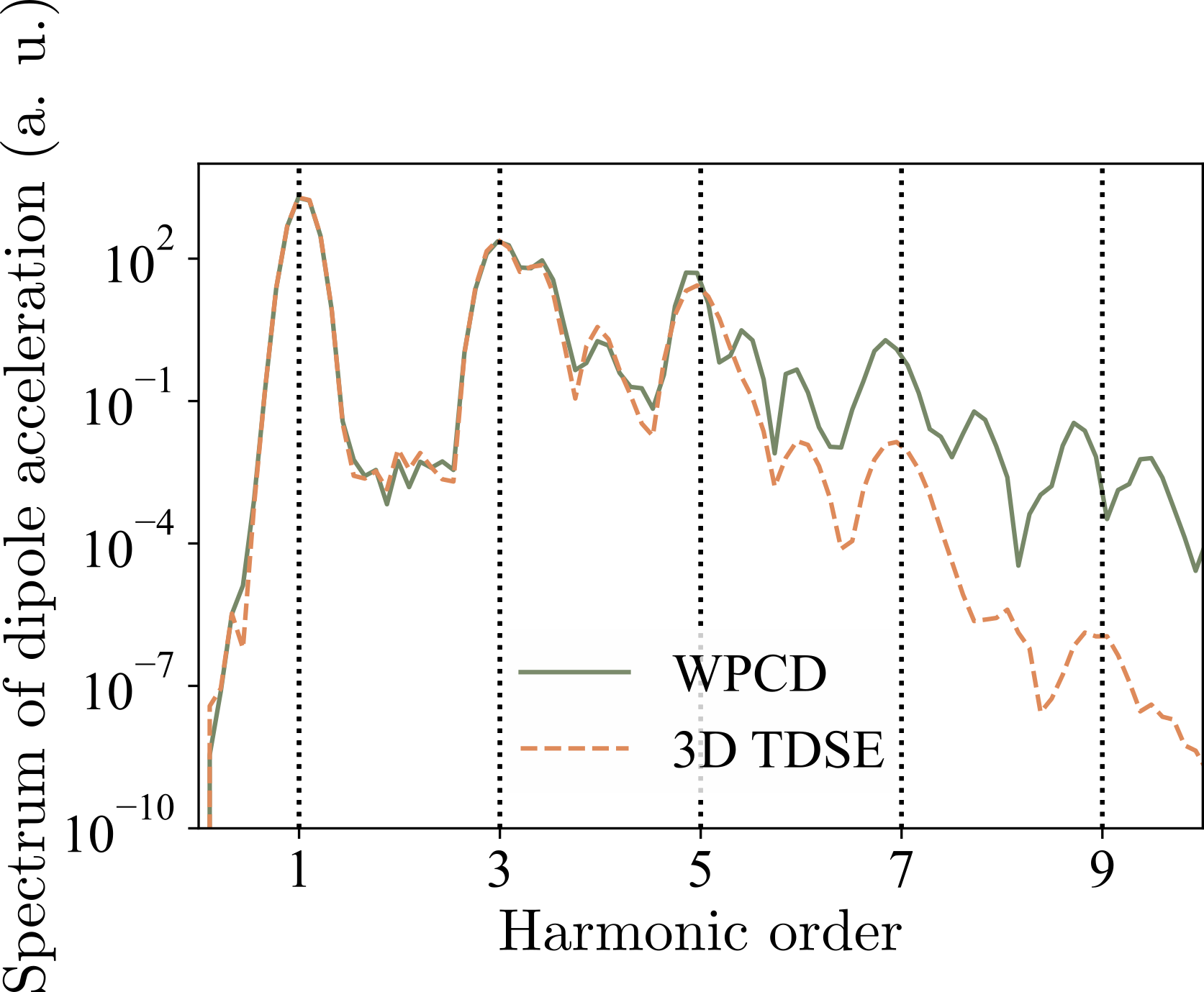}
    \caption{Power spectrum of the dipole acceleration for hydrogen driven by a 400~nm Gaussian pulse ($\mathcal{E}_0=0.02$\,a.u.\, $N_{\rm cycles}=10$), computed with the WPCD basis (solid green; $L_{\max}=6$, 60 continuum bins) and a state-of-the-art 3D TDSE calculation (dashed orange)~\cite{patchkovskii_simple_2016}. Vertical dotted lines mark odd harmonics.}
    \label{fig:spectrum_benchmark}
\end{figure}

To validate the WPCD basis before applying it to tracking control, we first compare the high-harmonic generation~(HHG) spectrum obtained within the 512-state basis against a state-of-the-art three-dimensional time-dependent Schr\"{o}dinger equation (TDSE) calculation~\cite{patchkovskii_simple_2016} for hydrogen driven by 400\,nm ($\omega~\approx 0.114$~a.u) Gaussian pulse with peak amplitude $\mathcal{E}_0=0.02$\,a.u.\ and $N_{\rm cycles}=10$. As shown in Fig.~\ref{fig:spectrum_benchmark} (source code~\cite{git_notebook_benchmark} for details), the WPCD calculation quantitatively reproduces the low-harmonic structure, with excellent agreement through approximately the fifth to seventh harmonic. The increasing deviations at higher harmonics are expected from the finite continuum discretization and the angular-momentum truncation ($L_{\max}=6$), which limit the representation of highly excited continuum dynamics~\cite{patchkovskii_simple_2016}. This agreement establishes that the WPCD basis accurately captures the response regime relevant to the present tracking-control study and constitutes, to our knowledge, its first application in strong-field optics.

\begin{figure}
    \centering
    \includegraphics[width=\linewidth]{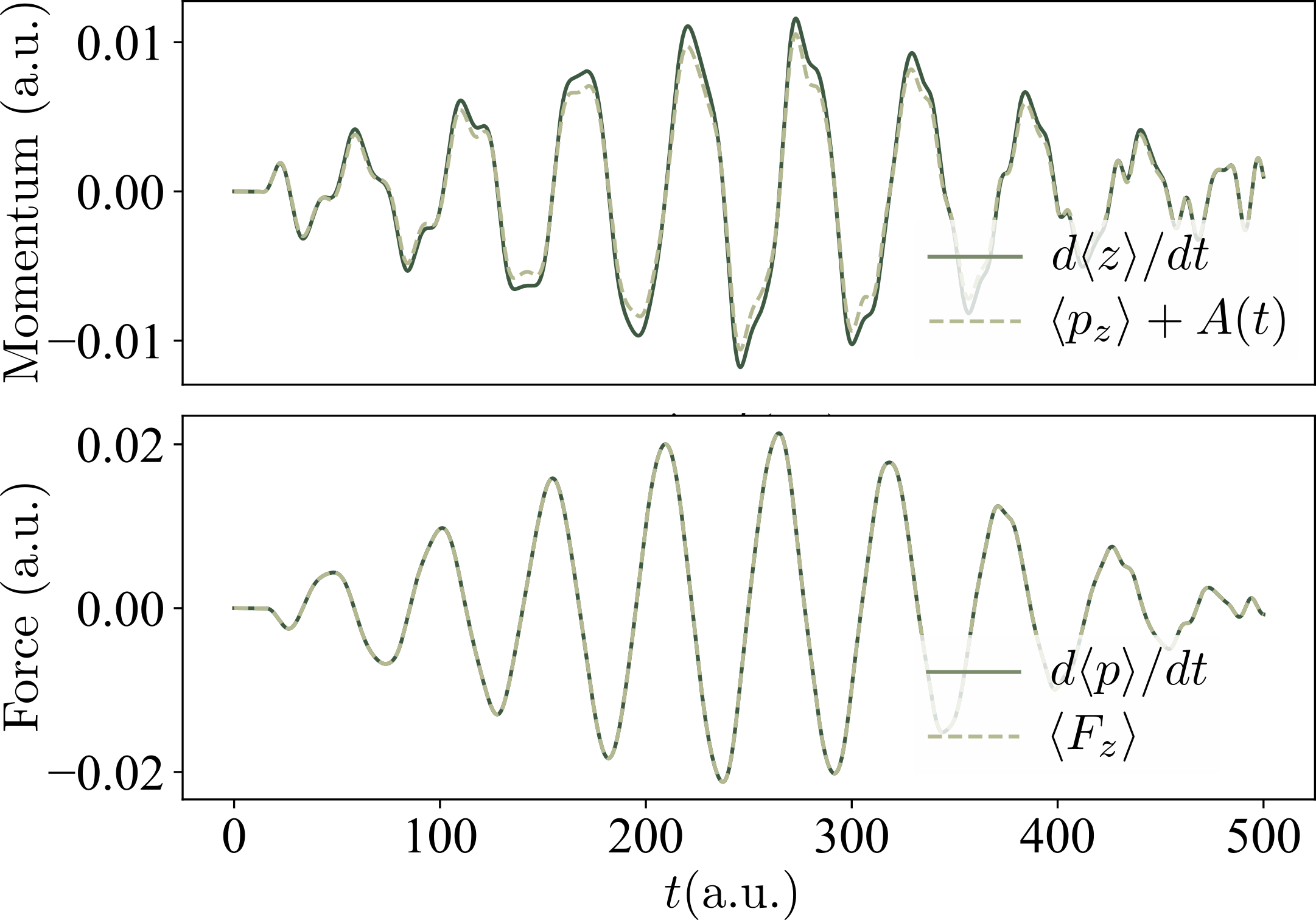}
    \caption{Numerical verification of the Ehrenfest theorems. The left- and right-hand sides of the two Ehrenfest relations, $\frac{d}{dt}\langle z\rangle = \langle p_z \rangle + A(t)$ (top) and $\frac{d}{dt}\langle p\rangle = \langle F_z \rangle$ (bottom), are computed independently from the simulated wave function and plotted as functions of time. The near-perfect overlap of the two curves in each panel indicates that the finite-dimensional representation used here introduces only a negligible violation of the Ehrenfest theorems, confirming the physical reliability of the simulation. See Ref.~\cite{git_notebook_benchmark} for computational details.}
    \label{fig:ehrenfest_theorems}
\end{figure}

As a final, and particularly stringent, benchmark, we numerically verify the Ehrenfest theorems~\cite{ehrenfest1927bemerkung, ballentine1994inadequacy}. It has been shown~\cite{bondar_conceptual_2013} that the Ehrenfest theorems cannot be satisfied exactly whenever finite-dimensional approximations of the coordinate and momentum operators are employed; in practice, they can therefore only hold approximately in numerical calculations. The physical reliability of a quantum simulation can thus be quantified by the degree to which the Ehrenfest theorems are violated: a physically meaningful calculation should minimize the discrepancy between the left- and right-hand sides of these identities. As shown in Fig.~\ref{fig:ehrenfest_theorems} (further details are given in Ref.~\cite{git_notebook_benchmark}), both Ehrenfest relations are satisfied. The small discrepancy visible in Fig.~\ref{fig:ehrenfest_theorems} can be further reduced by using a lower intensity of the driving field. Note that the preceding benchmark against the method of Ref.~\cite{patchkovskii_simple_2016} is formulated in the velocity gauge, and we employ the Ehrenfest theorems in the same gauge here for consistency.

Before tracking begins, the initial state must be prepared to seed the algorithm. The hydrogen ground state is parity-symmetric, so $\langle \Psi_0 | \hat{F}_z | \Psi_0 \rangle = 0$; inserting this into the tracking equation~\eqref{eq:tracking} yields $\mathcal{E}(t)=0$ at every subsequent time step, trapping the system at a fixed point. To break this symmetry, we first expose the atom to a short pump pulse (300 a.u.) with a $\sin^2$ envelope, peak amplitude $E_0 = 0.05$~a.u., carrier frequency $\omega~\approx 0.114$~a.u., and the number of cycles $N_{\rm cycles}=5$. This excites the atom into a superposition of stationary states while leaving the ground state dominant, producing a state with $\langle F_z \rangle \neq 0$ from which the self-consistent tracking algorithm can proceed. The magnitude of $\langle F_z\rangle$ after this preparatory pump sets the dynamic range available to tracking.

\begin{figure}
    \centering
    \includegraphics[width=0.9\linewidth]{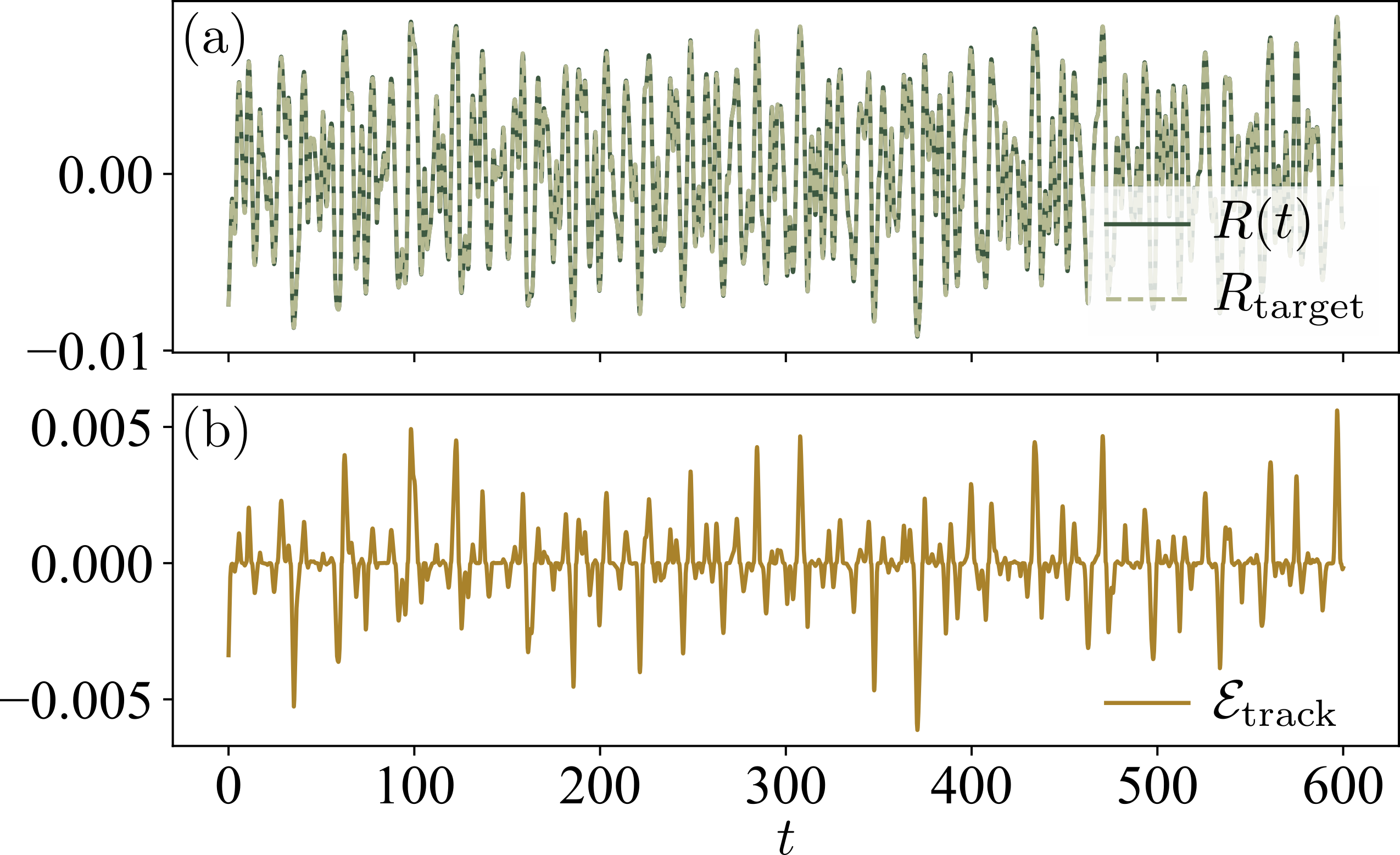}
    \caption{Tracking control for a cube-root response with $\nu = 3$, $\alpha = 0.03$ [Eq.~\eqref{eq:target_response}]. (a)~Dipole acceleration $d\langle p_z\rangle/dt$ (solid) and the prescribed target
    $R_{\rm target}(t) = 0.03\,\mathcal{E}^{1/3}$ (dashed). (b)~The tracking field $\mathcal{E}_{\rm track}(t)$ produced by the algorithm.}
    \label{fig:tracking_field}
\end{figure}

To demonstrate tracking control, we choose $\nu = 3$ and $\alpha = 0.03$ in Eq.~\eqref{eq:target_response}, targeting a cube-root response $R_{\rm target}(t) = 0.03\,\mathcal{E}^{1/3}$. Figure~\ref{fig:tracking_field}(a) compares the achieved response $R(t)$, i.e., a dipole acceleration $d{\langle p_z \rangle}/dt$ obtained from the evolved state, with the prescribed target response $R_{\rm target}(t)$. As seen, the two curves coincide to numerical precision throughout the entire propagation window, confirming that the tracking equation~\eqref{eq:tracking} is satisfied at every time step. Notably, this agreement persists through the zero crossings of $\mathcal{E}(t)$, where the target function $\operatorname{sgn}(\mathcal{E})\,|\mathcal{E}|^{1/\nu}$ has a divergent derivative and the tracking equation becomes ill-conditioned for standard root-finding methods. We avoid this singularity by solving instead for the auxiliary variable $u = \operatorname{sgn}(\mathcal{E})\,|\mathcal{E}|^{1/\nu}$, for which the tracking equation is smooth everywhere; $\mathcal{E}$ is then recovered as
$\mathcal{E} = \operatorname{sgn}(u)\,|u|^{\nu}$.

Figure~\ref{fig:tracking_field}(b) displays the tracking field $\mathcal{E}_{\rm track}(t)$ that produces this response. We emphasize that $\mathcal{E}_{\rm track}(t)$ is not prescribed but is determined entirely by the self-consistent solution of Eq.~\eqref{eq:tracking}. As demonstrated, it exhibits a complex temporal behavior, reflecting the nonlinear coupling between the field and the evolving quantum state.

\begin{figure}
    \centering
    \includegraphics[width=0.95\linewidth]{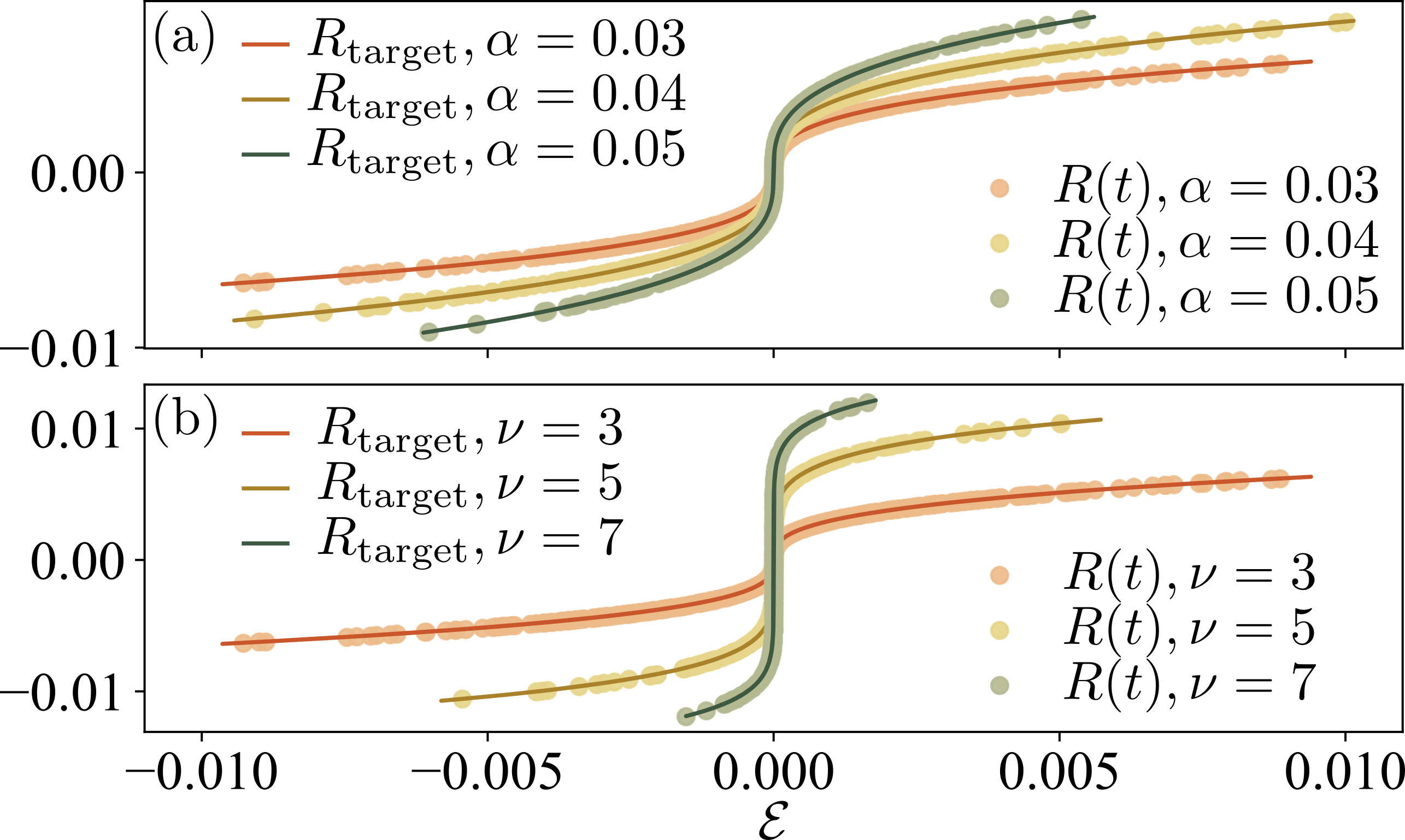}
    \caption{Achieved response $R$ versus tracking field
    $\mathcal{E}_{\rm track}$. Dots show the achieved response at each time step; solid curves show the target power law defined in Eq.~\eqref{eq:target_response}.
    (a)~Fixed $\nu = 3$, varying $\alpha = 0.03,\;0.04,\;0.05$. (b)~Fixed $\alpha = 0.03$, varying $\nu = 3, 5, 7$. All simulations use the same WPCD basis and
    initial-state preparation.}
    \label{fig:varying_params}
\end{figure}

To establish the universality of the proposed tracking algorithm and demonstrate that the sublinear power law in Eq.~\eqref{eq:target_response} can be engineered for a range of different values of $\alpha$ and $\nu$, we eliminate the explicit time dependence and construct parametric plots of the achieved response $R$ versus the tracking field $\mathcal{E}_{\rm track}$. Hence, each time step of the simulation yields one point in the $(\mathcal{E}_{\rm track},\,R)$ plane, and the resulting arrangement of points is compared directly with the target curve described by Eq.~\eqref{eq:target_response}. Figure~\ref{fig:varying_params} demonstrates results from independent tracking-control simulations for several values of $\alpha$ [Fig.~\ref{fig:varying_params}(a)] and $\nu$ [Fig.~\ref{fig:varying_params}(b)], all starting from the same prepared initial state.  In every case the data collapse onto the prescribed power-law curve, with the distinct curvatures clearly resolved. 
Together, these results demonstrate universality: the same hydrogen atom, under appropriate tracking control, can be made to exhibit different sublinear responses.

%-----------------------------------------------------
%%%%% Conclusions %%%%%
%-----------------------------------------------------
\section{Conclusions}
We have demonstrated that tracking control can produce a sublinear optical response in hydrogen atom. In this regime, the radiated field follows a fractional power of the driving field and therefore lies beyond the perturbative framework of conventional nonlinear optics. The method accurately realizes prescribed sublinear power laws for several values of the sublinearity parameter $\nu$ and scaling constants $\alpha$ by determining the required driving field self-consistently. These calculations are enabled by the WPCD basis, which, to our knowledge, is applied here for the first time to strong-field optics and quantum control. By treating bound and continuum states on equal footing within a highly efficient representation, this approach provides orders-of-magnitude speedup over converged three-dimensional TDSE calculations while retaining the relevant atomic dynamics.

Beyond establishing sublinear optical response, our results open a route to designing unconventional light--matter interactions by prescribing the desired relation between the driving field and the emitted response. Fractional-power scaling can enhance the relative response to weak inputs while compressing stronger signals, suggesting potential applications in detection of weak optical signals and dynamic-range control. Moreover, non-integer optical responses can also emerge naturally in systems with engineered symmetry and electronic structure, as illustrated by the predicted half-integer harmonics in strained graphene~\cite{Ornigotti2021}; the tracking-control framework developed in this study offers a systematic way to analyze and optimize such responses.

\acknowledgments

This work was supported by Army Research Office (ARO) (grant W911NF-23-1-0288; program manager Dr.~James Joseph). The views and conclusions contained in this document are those of the authors and should not be interpreted as representing the official policies, either expressed or implied, of ARO, or the U.S. Government. The U.S. Government is authorized to reproduce and distribute reprints for Government purposes notwithstanding any copyright notation herein.

\section*{Code availability} 

All codes and data used in this study can be found in Ref.~\cite{git_sublinear_optics}. 

\bibliography{main}

\end{document}